%% file: RpA.tex
\documentclass[ALICE,manyauthors]{cernphprep}

\usepackage{graphicx, subfigure}
\usepackage{amssymb}
\usepackage{amsmath}
\usepackage{lineno}
\usepackage{cite}
\usepackage{mciteplus}
\usepackage{hyperref}
\usepackage{multirow}
\usepackage{xspace}
\usepackage{color}

\newcommand{\pp}         {pp\xspace}

\newcommand{\ppb}        {\mbox{p--Pb}\xspace}

\newcommand{\pbpb}       {\mbox{Pb--Pb}\xspace}

\newcommand{\rtsnn}      {\ensuremath{\sqrt{s_\mathrm{NN}}}\xspace}
\newcommand{\gevc}       {~\ensuremath{\mathrm{GeV}\!/c}\xspace}
\newcommand{\tev}        {~\ensuremath{\mathrm{TeV}}\xspace}
\newcommand{\pt}         {\ensuremath{p_\mathrm{T}}\xspace}

\newcommand{\etacms}     {\ensuremath{\eta_\mathrm{cms}}\xspace}

\begin{document}%
%\linenumbers

%%%%%%%%%%%%%%%  Title page %%%%%%%%%%%%%%%%%%%%%%%%
%
\begin{titlepage}
\PHnumber{2014-088}               
\PHdate{07 May 2014}              % required

\title{Transverse momentum dependence of inclusive primary charged-particle 
production in \mbox{p--Pb} collisions at $\mathbf{\sqrt{s_{\rm{NN}}} = 5.02~TeV}$}

\ShortTitle{$\pt$ dependence of charged-particle production in p--Pb collisions}   % appears on right page headers

\Collaboration{ALICE Collaboration%
         \thanks{See Appendix~\ref{app:collab} for the list of collaboration
                      members}}
\ShortAuthor{ALICE Collaboration} % appears on left page headers, do not change

\begin{abstract}

The transverse momentum ($p_{\mathrm T}$) distribution of primary 
charged particles is measured at midrapidity in minimum-bias %(non-single-diffractive)
\mbox{p--Pb} collisions at \mbox{$\sqrt{s_{\mathrm{NN}}}=5.02$~TeV} 
with the ALICE detector at the LHC
in the range \mbox{$0.15<p_{\mathrm T}<50$~GeV/$c$}.
The spectra are compared to the expectation based on binary collision scaling of 
particle production in pp collisions, leading to a nuclear modification factor 
consistent with unity for $p_{\mathrm T}$ larger than 2~GeV/$c$,
with a weak indication of a Cronin-like enhancement for \pt around 4 \gevc.
The measurement is compared to theoretical calculations and to data in Pb--Pb 
collisions at \mbox{$\sqrt{s_{\mathrm{NN}}}=2.76$~TeV}.
\end{abstract}
\end{titlepage}

Measurements of particle production in proton-nucleus collisions at high energies 
enable the study of fundamental properties of Quantum Chromodynamics (QCD) 
over a broad range of parton fractional momentum $x$ and parton densities
(see \cite{Salgado:2011wc} for a review). 
They also provide reference measurements for the studies of deconfined matter 
created in nucleus-nucleus collisions \cite{Muller:2012zq}.

The first measurements of charged-particle production in minimum-bias \ppb collisions 
at the LHC at a centre-of-mass energy per nucleon-nucleon pair of
\rtsnn = 5.02\tev  \cite{ALICE:2012xs,ALICE:2012mj} 
showed that: i) the charged particle multiplicity density at midrapidity 
scales approximately with the number of participating nucleons 
($\langle N_{\mathrm{part}}\rangle=7.9\pm 0.6$ for minimum-bias collisions)
calculated in a Glauber model \cite{Miller:2007ri} and
ii) the transverse momentum (\pt) spectrum, measured in the range 
0.5--20\gevc \cite{ALICE:2012mj}, exhibits binary collision scaling above a few \gevc, 
as expected in the absence of any significant nuclear modification effect.
The latter is quantified by the nuclear modification factor, $R_{\mathrm{pPb}}$, the 
ratio of the \pt spectrum in \ppb collisions and a reference obtained by scaling 
the measurement in pp collisions with the number of binary nucleon-nucleon collisions 
in \ppb.
The preliminary result by the CMS collaboration \cite{CMS-PAS-HIN-12-017}
hints at an enhancement of particle production in p--Pb collisions above binary 
collision scaling, leading to $R_{\mathrm{pPb}}>1$, for \pt exceeding about 30\gevc.
The preliminary result by the ATLAS collaboration \cite{ATLAS-CONF-2014-029}
exhibits also, for collisions corresponding to 0--90\% centrality, $R_{\mathrm{pPb}}$ 
values above unity for \pt in the range 20--100\gevc. 

In this letter we present an update of our previously published \pt spectra 
of primary charged particles \cite{ALICE:2012mj} based on the 60 times 
larger sample size collected with the ALICE detector \cite{Abelev:2014ffa} in 2013 
in minimum-bias collisions. 
These data allow a significant extension of the transverse momentum range. 
The present analysis is essentially identical to the previous and therefore we update 
only the information related to the enlarged data set; the reader is referred to the 
earlier publications \cite{Abelev:2012eq,ALICE:2012mj,Abelev:2013ala,Abelev:2013bla} 
for a more detailed and complete description.

\begin{table}[h]
\centering
\label{tab_systerr} 
\begin{tabular}{ll}
Uncertainty & Value \\ \hline
Event selection & 0.6\% \\ 
Track selection & 1.0--5.5\% \\ 
Tracking efficiency & 3.0\% \\ 
$p_{\mathrm T}$ resolution & 0--1.3\% \\
$p_{\mathrm T}$ scale & 0--1.5\%  \\ 
Particle composition & 0.1--0.4\% \\ 
MC generator used for correction & 1.0\% \\
Secondary particle rejection & 0.5--4.0\% \\ 
Material budget & 0.2--1.5\% \\ 

\hline
Total for p--Pb, $p_{\mathrm T}$-dependent & 3.4--6.7\% \\
\hline
Normalization p--Pb & 3.1\% \\
\hline
\hline
Total for pp, $p_{\mathrm T}$-dependent  & 6.8--8.2\% \\ 
Normalization pp & 3.6\% \\
Nuclear overlap $\langle T_{\mathrm{pPb}} \rangle$  & 3.6\% \\
\end{tabular}
\caption{Systematic uncertainties on the $p_{\mathrm T}$-differential yields 
in p--Pb and pp collisions. % for $|\eta_{\mathrm{cms}}|<0.3$. 
The quoted ranges span the $p_{\mathrm T}$ dependence of the uncertainties
in the measured range, 0.15--50\gevc. Normalization uncertainties are also quoted.}
\end{table}

The ALICE minimum-bias trigger is defined by a coincidence of signals 
in detectors covering in pseudorapidity\footnote{In the laboratory frame 
\begin{math} \eta = -{\rm{ln}}[{\rm{tan}} (\vartheta/2)]\end{math}, 
with $\vartheta$ the polar angle between the charged particle
and the beam axis; the proton beam has negative $\eta$.}
$2.8<\eta<5.1$~(VZERO-A) and $-3.7<\eta<-1.7$~(VZERO-C).
In the 2013 data sample, 106 million events 
(corresponding to an integrated luminosity of 50.7$\pm$1.6 $\mu$b$^{-1}$)
satisfy the trigger and offline event-selection criteria, which select essentially 
non-single-diffractive (NSD) minimum-bias collisions.
The centre-of-mass pseudorapidity is defined as
$\eta_{\mathrm{cms}} = -\eta - |y_{\mathrm{NN}}|$, with the proton beam at positive rapidity;
$|y_{\mathrm{NN}}|=0.465$ is the rapidity of the centre-of-mass for nucleon-nucleon
collisions.
This equation is exact only for massless or very high \pt particles.
The spectra are corrected on a statistical basis using the measurements by ALICE 
in p--Pb collisions of the $\eta$ distribution of inclusive charged particle 
production \cite{ALICE:2012xs} and of the pion, kaon, and proton yields 
\cite{Abelev:2013haa}; this correction depends on the \etacms range and on \pt, 
reaching about 20\% for the lowest \pt bin. The systematic uncertainty of 
the particle composition \cite{Abelev:2013haa} leads to a systematic uncertainty
in our spectra of up to 0.4\%.

The systematic uncertainties on the spectra are evaluated as in previous analyses of 
\pp \cite{Abelev:2013ala}, \pbpb \cite{Abelev:2012eq}, and 
\ppb \cite{ALICE:2012mj} data.
The uncertainty due to the \pt scale is negligible below 20\gevc and reaches 1.5\% 
at 50\gevc.
The main contributions and the total uncertainties are listed in
Table~\ref{tab_systerr}. 

\begin{figure}[htb]
\begin{center}
\includegraphics[width=0.64\textwidth]{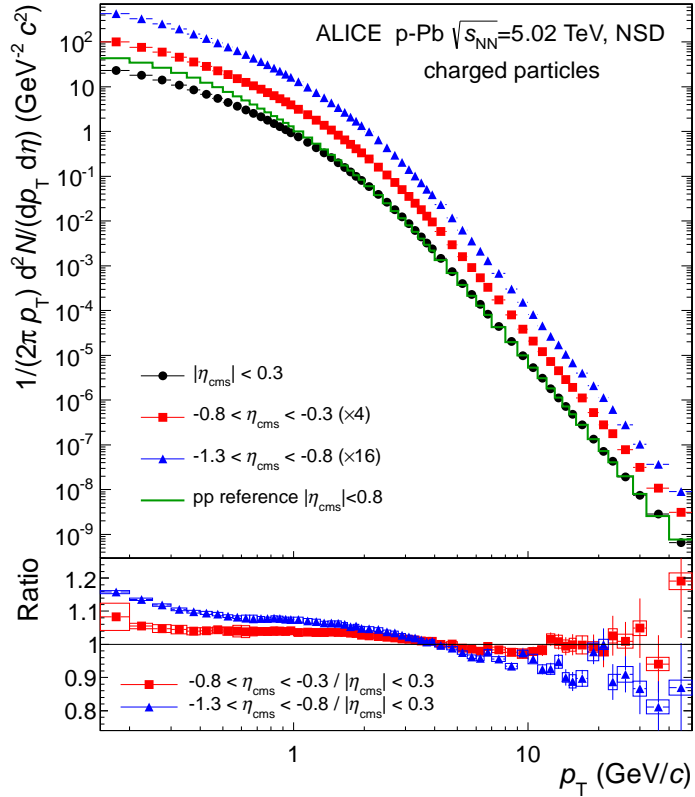}
\end{center}
\caption{Transverse momentum distributions of charged particles 
in minimum-bias (NSD) \mbox{p--Pb} collisions for different pseudorapidity 
ranges (upper panel).
The spectra are  scaled  by the factors indicated.  
The histogram represents the reference spectrum 
(cross section scaled by the nuclear overlap function, $T_{\mathrm{pPb}}$)
in inelastic pp collisions, 
determined in $|\eta|<0.8$.
The lower panel shows the ratio of spectra in \ppb at backward pseudorapidities
to that at $|\eta_{\mathrm{cms}}|<0.3$. 
The vertical bars (boxes) represent the statistical (systematic) uncertainties.
}
\label{fig_pt} 
\end{figure}

The  $p_{\mathrm T}$ spectra of charged particles measured in minimum-bias (NSD) 
\mbox{p--Pb} collisions at \mbox{$\sqrt{s_{\rm{NN}}} = 5.02$~TeV} are shown in 
Fig.~\ref{fig_pt} for the ranges $|\eta_{\mathrm{cms}}|<0.3$, 
$-0.8<\eta_{\mathrm{cms}}<-0.3$, and $-1.3<\eta_{\mathrm{cms}}<-0.8$.
The pp reference spectrum, ${\langle T_{\mathrm{pPb}} \rangle (1/2\pi\pt)
{\mathrm d}^{2}\sigma^{\mathrm{pp}}/{\mathrm d}\eta {\mathrm d}p_{\mathrm T} }$,
is also included. $\langle T_{\mathrm{pPb}}\rangle$ is the average nuclear overlap 
function, calculated using the Glauber model \cite{Alver:2008aq}, which gives 
$\langle T_{\mathrm{pPb}}\rangle=\langle N_{\mathrm{coll}}\rangle / \sigma_{\mathrm{NN}}
=0.0983 \pm 0.0035$ mb$^{-1}$, with $\langle N_{\mathrm{coll}}\rangle=6.9\pm 0.6$ 
and $\sigma_{\mathrm{NN}}=70\pm 5$ mb. 
Since the data in pp collisions \cite{Abelev:2013ala} indicate only a very small 
$\eta$ dependence of the \pt spectrum in the range measured by ALICE ($|\eta|<0.8$), 
our current reference spectrum is, differently than 
in \cite{Abelev:2013ala,ALICE:2012mj}, for $|\eta|<0.8$.
It was obtained by data interpolation at low \pt and by scaling the measurement
at  \mbox{$\sqrt{s} = 7$~TeV} with the ratio of spectra calculated with NLO pQCD 
at $\sqrt{s} = 5.02$ and 7~TeV \cite{Abelev:2013ala}.

In the lower panel of Fig.~\ref{fig_pt} the ratios of the spectra for 
backward ($-0.8<\eta_{\mathrm{cms}}<-0.3$ and $-1.3<\eta_{\mathrm{cms}}<-0.8$)
pseudorapidity ranges to that at  $|\eta_{\mathrm{cms}}|<0.3$ are shown.
The indication of a slight softening of the $p_{\mathrm T}$ spectrum when going from 
central to backward (Pb-side) pseudorapidity, observed already in the pilot-run 
data of 2012 \cite{ALICE:2012mj} 
(note opposite \etacms sign convention in \cite{ALICE:2012mj})
is confirmed with better significance and extended in $p_{\mathrm T}$ down to 0.15\gevc.

A good description of our earlier measurement of spectra in p--Pb collisions 
\cite{ALICE:2012mj} was achieved in the EPOS3 model \cite{Werner:2013tya} including 
a hydrodynamical description of the collision, while the PHSD model 
\cite{Konchakovski:2014wqa} significantly underestimated the spectra for \pt values 
of several \gevc.

\begin{figure}[htb]
\begin{center}
\includegraphics[width=0.56\textwidth]{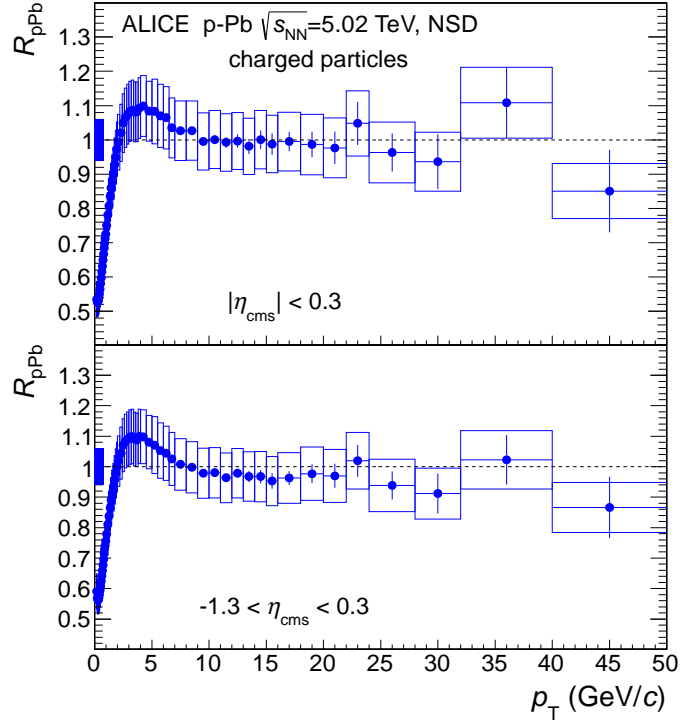}
\end{center}
\caption{The nuclear modification factor of charged particles as a function 
of transverse momentum, measured in minimum-bias (NSD) \mbox{p--Pb} collisions at 
$\sqrt{s_{\mathrm{NN}}} = 5.02$~TeV in two pseudorapidity ranges,
\mbox{$|\eta_{\mathrm{cms}}|<0.3$} and \mbox{$-1.3<\eta_{\mathrm{cms}}<0.3$}.
The statistical errors are represented by vertical bars, the systematic 
errors by boxes around data points. 
The relative systematic uncertainties on the normalization are shown as boxes 
around unity near $p_{\mathrm{T}}=0$.
}
\label{fig_rpa1} 
\end{figure}

In order to quantify nuclear effects in p--Pb collisions, the 
$p_{\mathrm T}$-differential yield relative to the pp reference, the nuclear 
modification factor, is calculated as:
\begin{equation}\label{eq_RpPb}
R_{\mathrm{pPb}} (p_{\mathrm T} ) = \frac{{\mathrm d}^{2}N^{\mathrm{pPb}}/
{\mathrm d}\eta{\mathrm d}p_{\mathrm{T}} }{\langle T_{\mathrm{pPb}} \rangle 
{\mathrm d}^{2}\sigma^{\mathrm{pp}}/{\mathrm d}\eta {\mathrm d}p_{\mathrm T} },
\end{equation}
where $N^{\mathrm{pPb}}$ is the charged particle yield in p--Pb 
collisions. 

The measurement of the nuclear modification factor $R_{\mathrm{pPb}}$ for 
charged particle production in \mbox{$|\eta_{\mathrm{cms}}|<0.3$} and
\mbox{$-1.3<\eta_{\mathrm{cms}}<0.3$} is shown in Fig.~\ref{fig_rpa1}. 
The uncertainties of the p--Pb and pp spectra are added in quadrature,
separately for the statistical and systematic uncertainties.
The systematic uncertainties are largely correlated between adjacent \pt bins.
The total systematic uncertainty on the normalization, the quadratic sum of 
the uncertainty on $\langle T_{\mathrm{pPb}} \rangle$, the normalization of the 
pp reference spectrum and the normalization of the p--Pb data, amounts to 6.0\%.
The $R_{\mathrm{pPb}}$ factor is consistent with unity up to \mbox{\pt=50\gevc}.
The average values of $R_{\mathrm{pPb}}$ in \mbox{$|\eta_{\mathrm{cms}}|<0.3$} are
0.995$\pm$0.007 (stat.)$\pm$0.084 (syst.)
for the \pt range 10--20\gevc, 
0.990$\pm$0.031 (stat.)$\pm$0.090 (syst.)
in the range 20--28\gevc and
0.969$\pm$0.056 (stat.)$\pm$0.090 (syst.) % corr. syst.
in the range 28--50\gevc. 
The systematic uncertainties are weighted averages of the values in \pt bins, 
with statistical uncertainties as inverse square weights; all values carry in 
addition the common overall normalization uncertainty of 6\%.

The data indicate a small enhancement, $R_{\mathrm{pPb}}$  above unity, 
barely significant within systematic errors, around 4\gevc, i.e. in the \pt 
region where the much stronger Cronin enhancement is seen at lower energies
\cite{PhysRevD.11.3105,Kopeliovich:2002yh}.

\begin{figure}[t]
\begin{center}
\includegraphics[width=0.65\textwidth]{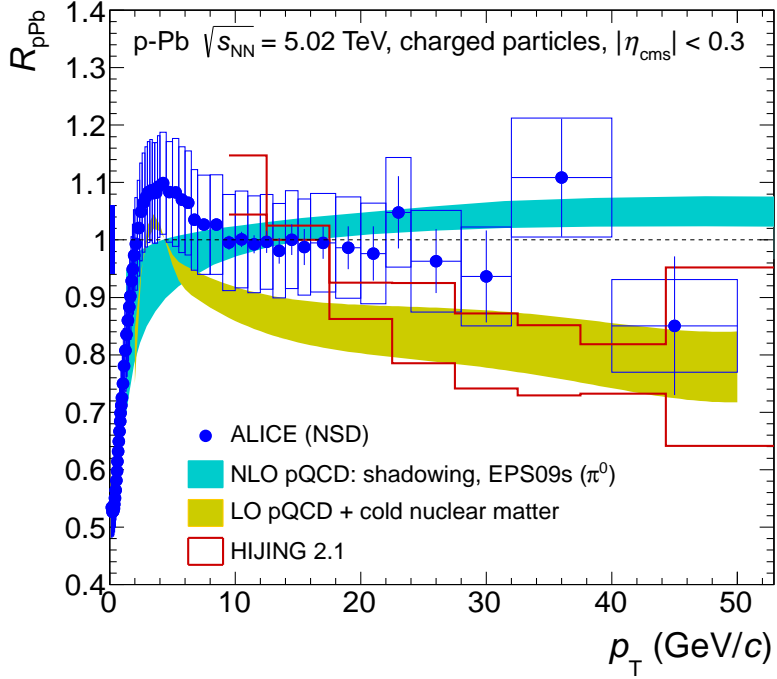}
\end{center}
\caption{Transverse momentum dependence of the nuclear modification 
factor $R_{\mathrm{pPb}}$ of charged particles measured in minimum-bias (NSD) 
\mbox{p--Pb} collisions at $\sqrt{s_{\mathrm{NN}}} = 5.02$~TeV. 
The ALICE data in $|\eta_{\mathrm{cms}}|<0.3$ (symbols) are compared to model 
calculations \cite{Helenius:2012wd,Kang:2012aa,Xu:2012au} (bands, see text 
for details).
The vertical bars (boxes) show the statistical (systematic) uncertainties.
The relative systematic uncertainty on the normalization is shown as a box 
around unity near $p_{\mathrm{T}}=0$. 
}
\label{fig_rpa2} 
\end{figure}

The p--Pb data provide important constraints to models of nuclear modification 
effects.
As an illustration, in Fig.~\ref{fig_rpa2} the measurement of $R_{\mathrm{pPb}}$ 
at \mbox{$|\eta_{\mathrm{cms}}|<0.3$} is compared to theoretical model predictions.
The predictions for shadowing \cite{Helenius:2012wd}, calculated at next-to-leading 
order (NLO) with the EPS09s nuclear modification of parton distribution functions 
describe the data for $\pt\gtrsim 6\gevc$.
The calculations are for $\pi^0$, which may explain the differences with respect 
to data at low \pt; for high \pt, the ALICE data on identified pions, kaons, and 
protons \cite{Abelev:2014laa} give support that the comparison of our data on 
inclusive charged particles to EPS09s calculations for $\pi^0$ is meaningful.
The LO pQCD model including cold nuclear matter effects \cite{Kang:2012aa}
exhibits a distinct trend of decreasing $R_{\mathrm{pPb}}$, which is not supported by 
the data.
The prediction with the HIJING 2.1 model, shown for two fragmentation schemes 
\cite{Xu:2012au}, exhibits a more pronounced trend of decreasing $R_{\mathrm{pPb}}$ 
at high \pt.
It is interesting to note that calculations with the EPOS LHC model \cite{Pierog:2013ria},
not included here, show a similar trend.
Several predictions based on the saturation (Color Glass Condensate) model are 
available \cite{Tribedy:2011aa,Albacete:2012xq,Rezaeian:2012ye};
they were shown previously \cite{ALICE:2012mj} to describe, 
in their range of validity, namely up to several \gevc, 
the $R_{\mathrm{pPb}}$ data.

\begin{figure}[thb]
\begin{center}
\includegraphics[width=0.65\textwidth]{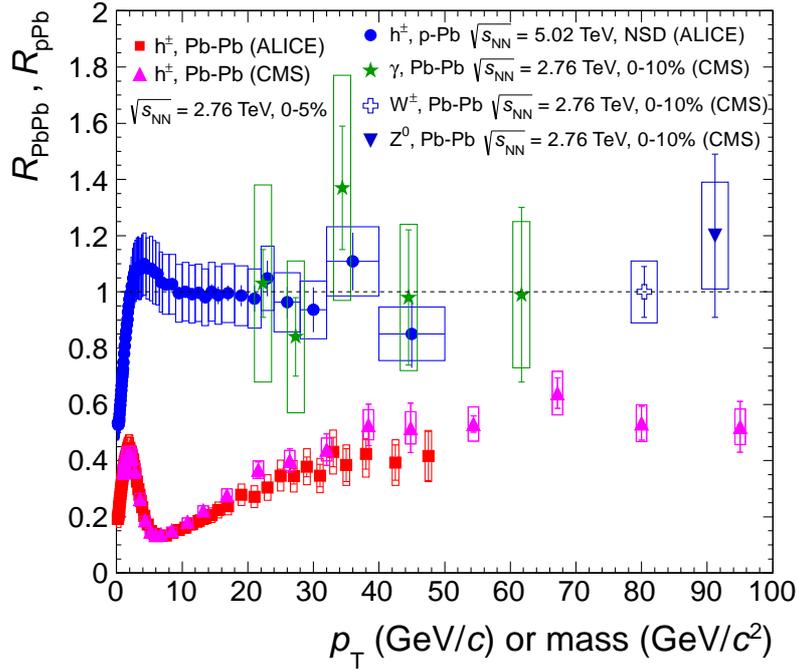}
\end{center}
\caption{Transverse momentum dependence of the nuclear modification 
factor $R_{\mathrm{pPb}}$ of charged particles (h$^\pm$) measured in minimum-bias (NSD) 
\mbox{p--Pb} collisions at $\sqrt{s_{\mathrm{NN}}} = 5.02$~TeV in comparison
to data on the nuclear modification factor $R_{\mathrm{PbPb}}$ in central 
Pb--Pb collisions at $\sqrt{s_{\mathrm{NN}}} = 2.76$~TeV.
The Pb--Pb data are for charged particle \cite{CMS:2012aa,Abelev:2012eq},
direct photon \cite{Chatrchyan:2012vq}, Z$^0$ \cite{Chatrchyan:2011ua}
and W$^{\pm}$ \cite{Chatrchyan:2012nt} production. All data are for midrapidity.
}
\label{fig_raa} 
\end{figure}

In Fig.~\ref{fig_raa} we compare the measurement of the nuclear modification 
factor for inclusive primary charged-particle ($h^{\pm}$) production in p--Pb 
collisions to that in central (0--5\% centrality) Pb--Pb collisions 
\cite{CMS:2012aa,Abelev:2012eq}.
The p--Pb data demonstrate that the suppression of hadron production at high \pt 
in Pb--Pb collisions, understood in theoretical models as a consequence of parton 
energy loss in (deconfined) QCD matter (see \cite{Abelev:2012eq} and references 
therein), has no contribution from initial state effects.
The ALICE p--Pb data show no sign of nuclear matter modification of
hadron production at high \pt and are therefore fully consistent with the 
observation of binary  collision scaling in Pb--Pb of observables which are not 
affected by hot QCD matter (direct photons \cite{Chatrchyan:2012vq} and vector 
bosons \cite{Chatrchyan:2011ua,Chatrchyan:2012nt}) 

In summary, we have extended our measurements of the charged-particle 
$p_{\mathrm T}$ spectra and nuclear modification factor in minimum-bias (NSD) 
\mbox{p--Pb} collisions at $\sqrt{s_{\mathrm{NN}}} = 5.02$~TeV.  
The results, covering a substantially-extended \pt range,  
$0.15<\pt<50\gevc$, exhibit, within uncertainties, no deviation from
binary collision scaling at high \pt;
the nuclear modification factor remains consistent with unity for $\pt\gtrsim 2\gevc$.
The data at high \pt are described by a prediction based on NLO pQCD 
calculations with PDF shadowing and further underline our earlier 
observation \cite{ALICE:2012mj} that initial state effects do not contribute to 
the strong suppression of hadron production at high \pt observed at the LHC in 
Pb--Pb collisions.

%%%%%%%% acknowledgements
\newenvironment{acknowledgement}{\relax}{\relax}
\begin{acknowledgement}
\section*{Acknowledgements}

We thank X.-N. Wang, K. Eskola, and I. Helenius for communications about their 
predictions.
\input{acknowledgements_march2013.tex}    %%%%%%% 
\end{acknowledgement}

\bibliography{RpA}
\newpage
%%%%%%%%% appendix with author list
\appendix
\section{The ALICE Collaboration}
\label{app:collab}
\input{Alice_Authorlist_2014-Apr-30-CERNPREP.tex}  %%%%%%% 

\end{document}

%% file: acknowledgements_march2013.tex
The ALICE Collaboration would like to thank all its engineers and technicians for their invaluable contributions to the construction of the experiment and the CERN accelerator teams for the outstanding performance of the LHC complex.
%\\
The ALICE Collaboration gratefully acknowledges the resources and support provided by all Grid centres and the Worldwide LHC Computing Grid (WLCG) collaboration.
%\\
The ALICE Collaboration acknowledges the following funding agencies for their support in building and
running the ALICE detector:
 %\\
State Committee of Science,  World Federation of Scientists (WFS)
and Swiss Fonds Kidagan, Armenia,
 %\\
Conselho Nacional de Desenvolvimento Cient\'{\i}fico e Tecnol\'{o}gico (CNPq), Financiadora de Estudos e Projetos (FINEP),
Funda\c{c}\~{a}o de Amparo \`{a} Pesquisa do Estado de S\~{a}o Paulo (FAPESP);
 %\\
National Natural Science Foundation of China (NSFC), the Chinese Ministry of Education (CMOE)
and the Ministry of Science and Technology of China (MSTC);
 %\\
Ministry of Education and Youth of the Czech Republic;
 %\\
Danish Natural Science Research Council, the Carlsberg Foundation and the Danish National Research Foundation;
 %\\
The European Research Council under the European Community's Seventh Framework Programme;
 %\\
Helsinki Institute of Physics and the Academy of Finland;
 %\\
French CNRS-IN2P3, the `Region Pays de Loire', `Region Alsace', `Region Auvergne' and CEA, France;
 %\\
German BMBF and the Helmholtz Association;
%\\
General Secretariat for Research and Technology, Ministry of
Development, Greece;
%\\
Hungarian OTKA and National Office for Research and Technology (NKTH);
 %\\
Department of Atomic Energy and Department of Science and Technology of the Government of India;
 %\\
Istituto Nazionale di Fisica Nucleare (INFN) and Centro Fermi -
Museo Storico della Fisica e Centro Studi e Ricerche "Enrico
Fermi", Italy;
 %\\
MEXT Grant-in-Aid for Specially Promoted Research, Ja\-pan;
 %\\
Joint Institute for Nuclear Research, Dubna;
 %\\
%Korea Foundation for International Cooperation of Science and Technology (KICOS);
National Research Foundation of Korea (NRF);
 %\\
CONACYT, DGAPA, M\'{e}xico, ALFA-EC and the EPLANET Program
(European Particle Physics Latin American Network)
 %\\
Stichting voor Fundamenteel Onderzoek der Materie (FOM) and the Nederlandse Organisatie voor Wetenschappelijk Onderzoek (NWO), Netherlands;
 %\\
Research Council of Norway (NFR);
 %\\
Polish Ministry of Science and Higher Education;
 %\\
National Science Centre, Poland;
 %\\
 Ministry of National Education/Institute for Atomic Physics and CNCS-UEFISCDI - Romania;
 %\\
Ministry of Education and Science of Russian Federation, Russian
Academy of Sciences, Russian Federal Agency of Atomic Energy,
Russian Federal Agency for Science and Innovations and The Russian
Foundation for Basic Research;
 %\\
Ministry of Education of Slovakia;
 %\\
Department of Science and Technology, South Africa;
 %\\
CIEMAT, EELA, Ministerio de Econom\'{i}a y Competitividad (MINECO) of Spain, Xunta de Galicia (Conseller\'{\i}a de Educaci\'{o}n),
CEA\-DEN, Cubaenerg\'{\i}a, Cuba, and IAEA (International Atomic Energy Agency);
 %\\
Swedish Research Council (VR) and Knut $\&$ Alice Wallenberg
Foundation (KAW);
 %\\
Ukraine Ministry of Education and Science;
 %\\
United Kingdom Science and Technology Facilities Council (STFC);
 %\\
The United States Department of Energy, the United States National
Science Foundation, the State of Texas, and the State of Ohio.

%% file: Alice_Authorlist_2014-Apr-30-CERNPREP.tex
% Collaboration: CERN-LHC-ALICE
% Generation Date is 2014/Apr/30

% How to use:
%%%%%%%%% appendix with author list
%\appendix
%\section{The ALICE Collaboration}
%\label{app:collab}
%\input{authors-list.tex}  %%%%%%% get the latest version before submitting

\begingroup
\small
\begin{flushleft}
B.~Abelev\Irefn{org69}\And
J.~Adam\Irefn{org37}\And
D.~Adamov\'{a}\Irefn{org77}\And
M.M.~Aggarwal\Irefn{org81}\And
M.~Agnello\Irefn{org105}\textsuperscript{,}\Irefn{org88}\And
A.~Agostinelli\Irefn{org26}\And
N.~Agrawal\Irefn{org44}\And
Z.~Ahammed\Irefn{org124}\And
N.~Ahmad\Irefn{org18}\And
I.~Ahmed\Irefn{org15}
% ...
\renewcommand\labelenumi{\textsuperscript{\theenumi}~}

\section*{Affiliation notes}
\renewcommand\theenumi{\roman{enumi}}
\begin{Authlist}
\item \Adef{0}Deceased
\item \Adef{idp14229584}{Also at: St. Petersburg State Polytechnical University}
\item \Adef{idp16132032}{Also at: Department of Applied Physics, Aligarh Muslim University, Aligarh, India}
\item \Adef{idp16814928}{Also at: M.V. Lomonosov Moscow State University, D.V. Skobeltsyn Institute of Nuclear Physics, Moscow, Russia}
\item \Adef{idp17058528}{Also at: University of Belgrade, Faculty of Physics and "Vin\v{c}a" Institute of Nuclear Sciences, Belgrade, Serbia}
\item \Adef{idp17364080}{Permanent Address: Permanent Address: Konkuk University, Seoul, Korea}
\item \Adef{idp17904416}{Also at: Institute of Theoretical Physics, University of Wroclaw, Wroclaw, Poland}
\item \Adef{idp18821776}{Also at: University of Kansas, Lawrence, KS, United States}
\end{Authlist}

\section*{Collaboration Institutes}
\renewcommand\theenumi{\arabic{enumi}~}
\begin{Authlist}

\item \Idef{org1}A.I. Alikhanyan National Science Laboratory (Yerevan Physics Institute) Foundation, Yerevan, Armenia
\item \Idef{org2}Benem\'{e}rita Universidad Aut\'{o}noma de Puebla, Puebla, Mexico
\item \Idef{org3}Bogolyubov Institute for Theoretical Physics, Kiev, Ukraine
\item \Idef{org4}Bose Institute, Department of Physics and Centre for Astroparticle Physics and Space Science (CAPSS), Kolkata, India
\item \Idef{org5}Budker Institute for Nuclear Physics, Novosibirsk, Russia
\item \Idef{org6}California Polytechnic State University, San Luis Obispo, CA, United States
\item \Idef{org7}Central China Normal University, Wuhan, China
\item \Idef{org8}Centre de Calcul de l'IN2P3, Villeurbanne, France
\item \Idef{org9}Centro de Aplicaciones Tecnol\'{o}gicas y Desarrollo Nuclear (CEADEN), Havana, Cuba
\item \Idef{org10}Centro de Investigaciones Energ\'{e}ticas Medioambientales y Tecnol\'{o}gicas (CIEMAT), Madrid, Spain
\item \Idef{org11}Centro de Investigaci\'{o}n y de Estudios Avanzados (CINVESTAV), Mexico City and M\'{e}rida, Mexico
\item \Idef{org12}Centro Fermi - Museo Storico della Fisica e Centro Studi e Ricerche ``Enrico Fermi'', Rome, Italy
\item \Idef{org13}Chicago State University, Chicago, USA
\item \Idef{org14}Commissariat \`{a} l'Energie Atomique, IRFU, Saclay, France
\item \Idef{org15}COMSATS Institute of Information Technology (CIIT), Islamabad, Pakistan
\item \Idef{org16}Departamento de F\'{\i}sica de Part\'{\i}culas and IGFAE, Universidad de Santiago de Compostela, Santiago de Compostela, Spain
\item \Idef{org17}Department of Physics and Technology, University of Bergen, Bergen, Norway
\item \Idef{org18}Department of Physics, Aligarh Muslim University, Aligarh, India
\item \Idef{org19}Department of Physics, Ohio State University, Columbus, OH, United States
\item \Idef{org20}Department of Physics, Sejong University, Seoul, South Korea
\item \Idef{org21}Department of Physics, University of Oslo, Oslo, Norway
\item \Idef{org22}Dipartimento di Fisica dell'Universit\`{a} 'La Sapienza' and Sezione INFN Rome, Italy
\item \Idef{org23}Dipartimento di Fisica dell'Universit\`{a} and Sezione INFN, Cagliari, Italy
\item \Idef{org24}Dipartimento di Fisica dell'Universit\`{a} and Sezione INFN, Trieste, Italy
\item \Idef{org25}Dipartimento di Fisica dell'Universit\`{a} and Sezione INFN, Turin, Italy
\item \Idef{org26}Dipartimento di Fisica e Astronomia dell'Universit\`{a} and Sezione INFN, Bologna, Italy
\item \Idef{org27}Dipartimento di Fisica e Astronomia dell'Universit\`{a} and Sezione INFN, Catania, Italy
\item \Idef{org28}Dipartimento di Fisica e Astronomia dell'Universit\`{a} and Sezione INFN, Padova, Italy
\item \Idef{org29}Dipartimento di Fisica `E.R.~Caianiello' dell'Universit\`{a} and Gruppo Collegato INFN, Salerno, Italy
\item \Idef{org30}Dipartimento di Scienze e Innovazione Tecnologica dell'Universit\`{a} del  Piemonte Orientale and Gruppo Collegato INFN, Alessandria, Italy
\item \Idef{org31}Dipartimento Interateneo di Fisica `M.~Merlin' and Sezione INFN, Bari, Italy
\item \Idef{org32}Division of Experimental High Energy Physics, University of Lund, Lund, Sweden
\item \Idef{org33}Eberhard Karls Universit\"{a}t T\"{u}bingen, T\"{u}bingen, Germany
\item \Idef{org34}European Organization for Nuclear Research (CERN), Geneva, Switzerland
\item \Idef{org35}Faculty of Engineering, Bergen University College, Bergen, Norway
\item \Idef{org36}Faculty of Mathematics, Physics and Informatics, Comenius University, Bratislava, Slovakia
\item \Idef{org37}Faculty of Nuclear Sciences and Physical Engineering, Czech Technical University in Prague, Prague, Czech Republic
\item \Idef{org38}Faculty of Science, P.J.~\v{S}af\'{a}rik University, Ko\v{s}ice, Slovakia
\item \Idef{org39}Frankfurt Institute for Advanced Studies, Johann Wolfgang Goethe-Universit\"{a}t Frankfurt, Frankfurt, Germany
\item \Idef{org40}Gangneung-Wonju National University, Gangneung, South Korea
\item \Idef{org41}Gauhati University, Department of Physics, Guwahati, India
\item \Idef{org42}Helsinki Institute of Physics (HIP), Helsinki, Finland
\item \Idef{org43}Hiroshima University, Hiroshima, Japan
\item \Idef{org44}Indian Institute of Technology Bombay (IIT), Mumbai, India
\item \Idef{org45}Indian Institute of Technology Indore, Indore (IITI), India
\item \Idef{org46}Institut de Physique Nucl\'eaire d'Orsay (IPNO), Universit\'e Paris-Sud, CNRS-IN2P3, Orsay, France
\item \Idef{org47}Institut f\"{u}r Informatik, Johann Wolfgang Goethe-Universit\"{a}t Frankfurt, Frankfurt, Germany
\item \Idef{org48}Institut f\"{u}r Kernphysik, Johann Wolfgang Goethe-Universit\"{a}t Frankfurt, Frankfurt, Germany
\item \Idef{org49}Institut f\"{u}r Kernphysik, Westf\"{a}lische Wilhelms-Universit\"{a}t M\"{u}nster, M\"{u}nster, Germany
\item \Idef{org50}Institut Pluridisciplinaire Hubert Curien (IPHC), Universit\'{e} de Strasbourg, CNRS-IN2P3, Strasbourg, France
\item \Idef{org51}Institute for Nuclear Research, Academy of Sciences, Moscow, Russia
\item \Idef{org52}Institute for Subatomic Physics of Utrecht University, Utrecht, Netherlands
\item \Idef{org53}Institute for Theoretical and Experimental Physics, Moscow, Russia
\item \Idef{org54}Institute of Experimental Physics, Slovak Academy of Sciences, Ko\v{s}ice, Slovakia
\item \Idef{org55}Institute of Physics, Academy of Sciences of the Czech Republic, Prague, Czech Republic
\item \Idef{org56}Institute of Physics, Bhubaneswar, India
\item \Idef{org57}Institute of Space Science (ISS), Bucharest, Romania
\item \Idef{org58}Instituto de Ciencias Nucleares, Universidad Nacional Aut\'{o}noma de M\'{e}xico, Mexico City, Mexico
\item \Idef{org59}Instituto de F\'{\i}sica, Universidad Nacional Aut\'{o}noma de M\'{e}xico, Mexico City, Mexico
\item \Idef{org60}iThemba LABS, National Research Foundation, Somerset West, South Africa
\item \Idef{org61}Joint Institute for Nuclear Research (JINR), Dubna, Russia
\item \Idef{org62}Korea Institute of Science and Technology Information, Daejeon, South Korea
\item \Idef{org63}KTO Karatay University, Konya, Turkey
\item \Idef{org64}Laboratoire de Physique Corpusculaire (LPC), Clermont Universit\'{e}, Universit\'{e} Blaise Pascal, CNRS--IN2P3, Clermont-Ferrand, France
\item \Idef{org65}Laboratoire de Physique Subatomique et de Cosmologie, Universit\'{e} Grenoble-Alpes, CNRS-IN2P3, Grenoble, France
\item \Idef{org66}Laboratori Nazionali di Frascati, INFN, Frascati, Italy
\item \Idef{org67}Laboratori Nazionali di Legnaro, INFN, Legnaro, Italy
\item \Idef{org68}Lawrence Berkeley National Laboratory, Berkeley, CA, United States
\item \Idef{org69}Lawrence Livermore National Laboratory, Livermore, CA, United States
\item \Idef{org70}Moscow Engineering Physics Institute, Moscow, Russia
\item \Idef{org71}National Centre for Nuclear Studies, Warsaw, Poland
\item \Idef{org72}National Institute for Physics and Nuclear Engineering, Bucharest, Romania
\item \Idef{org73}National Institute of Science Education and Research, Bhubaneswar, India
\item \Idef{org74}Niels Bohr Institute, University of Copenhagen, Copenhagen, Denmark
\item \Idef{org75}Nikhef, National Institute for Subatomic Physics, Amsterdam, Netherlands
\item \Idef{org76}Nuclear Physics Group, STFC Daresbury Laboratory, Daresbury, United Kingdom
\item \Idef{org77}Nuclear Physics Institute, Academy of Sciences of the Czech Republic, \v{R}e\v{z} u Prahy, Czech Republic
\item \Idef{org78}Oak Ridge National Laboratory, Oak Ridge, TN, United States
\item \Idef{org79}Petersburg Nuclear Physics Institute, Gatchina, Russia
\item \Idef{org80}Physics Department, Creighton University, Omaha, NE, United States
\item \Idef{org81}Physics Department, Panjab University, Chandigarh, India
\item \Idef{org82}Physics Department, University of Athens, Athens, Greece
\item \Idef{org83}Physics Department, University of Cape Town, Cape Town, South Africa
\item \Idef{org84}Physics Department, University of Jammu, Jammu, India
\item \Idef{org85}Physics Department, University of Rajasthan, Jaipur, India
\item \Idef{org86}Physik Department, Technische Universit\"{a}t M\"{u}nchen, Munich, Germany
\item \Idef{org87}Physikalisches Institut, Ruprecht-Karls-Universit\"{a}t Heidelberg, Heidelberg, Germany
\item \Idef{org88}Politecnico di Torino, Turin, Italy
\item \Idef{org89}Purdue University, West Lafayette, IN, United States
\item \Idef{org90}Pusan National University, Pusan, South Korea
\item \Idef{org91}Research Division and ExtreMe Matter Institute EMMI, GSI Helmholtzzentrum f\"ur Schwerionenforschung, Darmstadt, Germany
\item \Idef{org92}Rudjer Bo\v{s}kovi\'{c} Institute, Zagreb, Croatia
\item \Idef{org93}Russian Federal Nuclear Center (VNIIEF), Sarov, Russia
\item \Idef{org94}Russian Research Centre Kurchatov Institute, Moscow, Russia
\item \Idef{org95}Saha Institute of Nuclear Physics, Kolkata, India
\item \Idef{org96}School of Physics and Astronomy, University of Birmingham, Birmingham, United Kingdom
\item \Idef{org97}Secci\'{o}n F\'{\i}sica, Departamento de Ciencias, Pontificia Universidad Cat\'{o}lica del Per\'{u}, Lima, Peru
\item \Idef{org98}Sezione INFN, Bari, Italy
\item \Idef{org99}Sezione INFN, Bologna, Italy
\item \Idef{org100}Sezione INFN, Cagliari, Italy
\item \Idef{org101}Sezione INFN, Catania, Italy
\item \Idef{org102}Sezione INFN, Padova, Italy
\item \Idef{org103}Sezione INFN, Rome, Italy
\item \Idef{org104}Sezione INFN, Trieste, Italy
\item \Idef{org105}Sezione INFN, Turin, Italy
\item \Idef{org106}SSC IHEP of NRC Kurchatov institute, Protvino, Russia
\item \Idef{org107}SUBATECH, Ecole des Mines de Nantes, Universit\'{e} de Nantes, CNRS-IN2P3, Nantes, France
\item \Idef{org108}Suranaree University of Technology, Nakhon Ratchasima, Thailand
\item \Idef{org109}Technical University of Split FESB, Split, Croatia
\item \Idef{org110}The Henryk Niewodniczanski Institute of Nuclear Physics, Polish Academy of Sciences, Cracow, Poland
\item \Idef{org111}The University of Texas at Austin, Physics Department, Austin, TX, USA
\item \Idef{org112}Universidad Aut\'{o}noma de Sinaloa, Culiac\'{a}n, Mexico
\item \Idef{org113}Universidade de S\~{a}o Paulo (USP), S\~{a}o Paulo, Brazil
\item \Idef{org114}Universidade Estadual de Campinas (UNICAMP), Campinas, Brazil
\item \Idef{org115}University of Houston, Houston, TX, United States
\item \Idef{org116}University of Jyv\"{a}skyl\"{a}, Jyv\"{a}skyl\"{a}, Finland
\item \Idef{org117}University of Liverpool, Liverpool, United Kingdom
\item \Idef{org118}University of Tennessee, Knoxville, TN, United States
\item \Idef{org119}University of Tokyo, Tokyo, Japan
\item \Idef{org120}University of Tsukuba, Tsukuba, Japan
\item \Idef{org121}University of Zagreb, Zagreb, Croatia
\item \Idef{org122}Universit\'{e} de Lyon, Universit\'{e} Lyon 1, CNRS/IN2P3, IPN-Lyon, Villeurbanne, France
\item \Idef{org123}V.~Fock Institute for Physics, St. Petersburg State University, St. Petersburg, Russia
\item \Idef{org124}Variable Energy Cyclotron Centre, Kolkata, India
\item \Idef{org125}Vestfold University College, Tonsberg, Norway
\item \Idef{org126}Warsaw University of Technology, Warsaw, Poland
\item \Idef{org127}Wayne State University, Detroit, MI, United States
\item \Idef{org128}Wigner Research Centre for Physics, Hungarian Academy of Sciences, Budapest, Hungary
\item \Idef{org129}Yale University, New Haven, CT, United States
\item \Idef{org130}Yonsei University, Seoul, South Korea
\item \Idef{org131}Zentrum f\"{u}r Technologietransfer und Telekommunikation (ZTT), Fachhochschule Worms, Worms, Germany
\end{Authlist}
\endgroup